\documentclass[a4paper,twocolumn,11pt,accepted=2023-12-11]{quantumarticle}
\pdfoutput=1

\usepackage[utf8]{inputenc}
\usepackage{braket}

\usepackage{graphicx}
\usepackage{dcolumn}
\usepackage{bm}
\usepackage{hyperref}

\usepackage{balance}
\usepackage{multirow}

\usepackage{xcolor}

\usepackage{mathrsfs}

\begin{document}

\title{Pipelined correlated minimum weight perfect matching of the surface code}

\author{Alexandru Paler}
\affiliation{Aalto University, Espoo 02150, Finland}
\affiliation{University of Texas at Dallas, Richardson, TX 75080, USA}

\author{Austin G. Fowler}
\affiliation{Google Inc., Santa Barbara, 93117 CA, USA}

\begin{abstract}
We describe a pipeline approach to decoding the surface code using minimum weight perfect matching, including taking into account correlations between detection events. An independent no-communication parallelizable processing stage reweights the graph according to likely correlations, followed by another no-communication parallelizable stage for high confidence matching. A later general stage finishes the matching. This is a simplification of previous correlated matching techniques which required a complex interaction between general matching and re-weighting the graph. Despite this simplification, which gives correlated matching a better chance of achieving real-time processing, we find the logical error rate practically unchanged. We validate the new algorithm on the fully fault-tolerant toric, unrotated, and rotated surface codes, all with standard depolarizing noise. We expect these techniques to be applicable to a wide range of other decoders.
\end{abstract}

\maketitle

\section{Introduction}

The surface code \cite{Brav98,Denn02,Raus07,Raus07d,Fowl12f,Fowl18,Liti18b,Fowl19} is experimentally attractive as it requires only a 2D array of qubits with nearest neighbor interactions, and gate error rates of order 0.1\% to achieve post-classical fault-tolerant quantum computation with fewer than 1M qubits \cite{Kivl19}. Decoding the surface code must be done both accurately and in real time, and many algorithms have been proposed \cite{Andr12,Ducl13,Woot13,Woot13b,Fowl13g,Bair18, delfosse2021almost, demarti2023decoding}.

Realizing a scalable, accurate, real-time surface code decoder remains an open problem. To move closer to this goal, we improve the algorithm used in~\cite{Fowl13g} which performs a correlated-error version of minimum weight perfect matching~\cite{Edmo65a,Edmo65b}.

The method described by~\cite{Fowl13g} required multiple runs of the matching decoder in order to take correlations between errors into account. Minimum weight perfect matching decoders operate on graphs of weighted edges, and the graphs had to be locally reweighted according to correlations in the underlying error model~\cite{Fowl13g}. Reweighting triggered the erasure of partial solutions (matchings), leading to significant computational overheads -- the entire matching had to be recomputed from scratch. Each reweighting operation had the overhead of a usual matching decoder $\mathcal{O}(d^6\log (d))$~\cite{higgott2022pymatching} where $d$ is the distance of the surface code.

Our contribution is a pipelined method that significantly reduces the complexity and the execution time of the steps necessary for performing correlated decoding. Instead of re-running for an unknown number of times the full decoder of complexity $\mathcal{O}(d^6\log (d))$ in order to take reweights into account, we run a pre-matching decoder of complexity $\mathcal{O}(d^2)$ (Section~\ref{sec:appendix}). The first decoding pipeline stage is pre-matching for reweighting, the second (optional) stage is pre-matching on the reweighted graph, and the third stage is the usual decoder using the pre-matched outputs and the reweighted graph. Practically, this works shows that fast correlated rematching is possible. Moreover, we avoid the significant overhead while maintaining decoding performance.

In Section~\ref{sec:corr}, we describe how we extract and represent correlations in our circuits and error models. Section~\ref{sec:prem} describes a round by round algorithm to use the stream of detection events to infer correlations, reweight the detection graph, and perform a partial matching of detection events. This is followed by the algorithm of \cite{Fowl13g, demarti2023decoding}, with no further reweighting of the detection graph. Section~\ref{sec:simulations} describes circuit level depolarizing noise simulations of the toric, unrotated, and rotated surface codes, all with and without correlated graph reweighting. The Appendix describes our pre-matching method, motivates its design for almost perfect parallelism and a simple method for ensure the correctness of the distributed/parallel algorithm.

\section{Methods}

We perform correlated reweighting and an initial partial matching in a strictly round by round manner, with no backtracking or revision, and without sacrificing performance in terms of the achieved logical error rate.

\begin{figure}[!t]
\centering
    \includegraphics[width=\columnwidth]{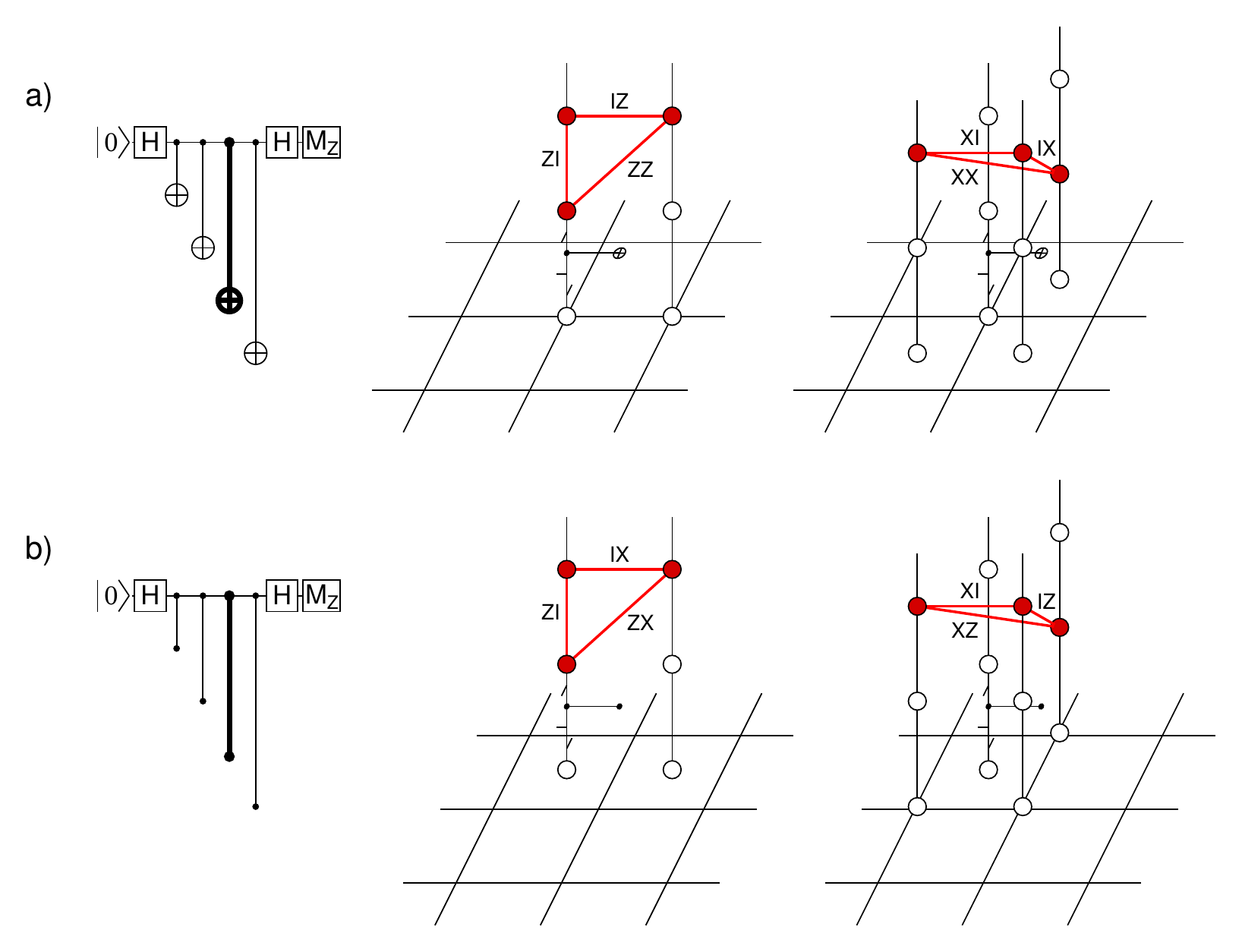}
    \caption{Three rounds of unrotated surface code: a) $X$ stabilizers measured on vertices; b) $Z$ stabilizers measured on faces. Time runs horizontally in the 2D and vertically in the 3D circuits. Bolded gates suffer all possible errors with only those leading to pairs of detection events (red) shown.}
    \label{fig:de_ex}
\end{figure}

Our methods, including decoding by matching, is performed on a graph called \emph{the detection graph}. The graph is generated by the analysis described in Section~\ref{sec:corr}, and the graph edges are located in space-time (e.g. Fig.~\ref{fig:de_ex}). Each edge has a probability $p$, and typically, it is more convenient to talk about the weight $w$ rather than their probability $p$.

Weights are associated to the edges in such a way that the most probable error corresponds to the minimum weight matching of the detection graph. Usually, most matching decoders use $w=-\ln \frac{p}{1-p}$ where $p$ is a probability of a single error associated with the qubit or gates of the code~\cite{Denn02}. Technically, the edge probability is the probability of an odd number of independent errors occurring. In this work, we continue using $w=-\ln p$ similarly to how it was used in \cite{Fowl13g}. This weight approximation is sufficient for usual values of $p$, avoids the issue of negative weights~\cite{higgott2022pymatching}, and achieves low logical error rates, as we shall see in Section~\ref{sec:simulations}.

\subsection{Correlated pre-analysis of the surface code}
\label{sec:corr}

To illustrate how an analysis of correlations in the surface code can be performed, consider a single two-qubit gate in multiple rounds of surface code stabilizer measurement. We assume depolarizing noise, but with each of the 15 nontrivial tensor products of $I$, $X$, $Y$, $Z$ potentially having a different probability. Each error on this gate can lead to anything from 0 to 4 detection events. Fig.~\ref{fig:de_ex} gives two examples of what can be observed.

Internally, for each gate we have a list of errors, and each error has a list of coordinates of detection events. Practically, our information is structured as follows \emph{Gate $\rightarrow$ Errors $\rightarrow$ DetectionEventCoords}. Processing this information begins with identifying those errors that lead to: a) single detection events; b) pairs of detection events; c) 2+ detection events.

First, we focus on single detection events. These are represented graphically by an edge to an unspecified boundary. The coordinate of the generating detection event uniquely identifies each boundary edge. A single gate may generate no or many boundary edges. Each boundary edge keeps a list of errors that generated it, and each error on this gate that generates a boundary edge is appended to the appropriate error list.

Second, we focus on errors that lead to pairs of detection event. If both detection events are associated with boundary edges, we skip that error for the moment. Otherwise, we associate an edge with the detection event coordinates. As before, such an edge keeps a list of errors that generated it, and we append each generating error to the appropriate edge's error list.

Finally, we treat the 2+ detection events. The edges found so far form a basis, meaning all remaining errors generating 2+ detection events can be uniquely decomposed into two or more edges from this basis. Such errors are added to the error lists associated with each decomposed edge, and each copy of this error will be specially annotated with the list of decomposed edges for later processing.

\begin{figure}[!t]
    \centering
    \includegraphics[width=\columnwidth]{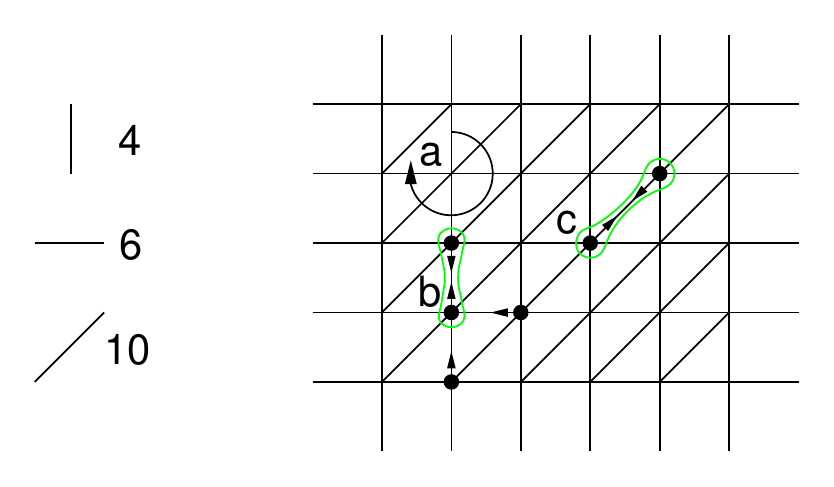}
    \caption{Pre-matching. Example of a graph with vertical edge weights 4, horizontal edge weights 6, and diagonal edge weights 8. a) Arbitrarily chosen ordering of the edges emanating from a vertex. b) Detection event with two lowest weight neighboring detection events, namely those vertically above and below it. Given the arbitrary ordering, the one above will be chosen. Note that since the detection event above has only a single neighbor, this choice will be mutual, indicated by a green bubble. c) Detection event with two equal lowest weight diagonal neighbors, the mutual chosen pair is shown.}
\label{fig:prematching}
\end{figure}

Some of the errors associated with each edge will have a list of decomposed edges. These decomposed edges, which were obtained from the 2+ detection events, form the basis of the correlated analysis. In the graph we have the information structured like \emph{ParentEdge $\leftarrow$ Error $\leftarrow$ DecomposedEdges}. The decomposed edges that are different from the parent are called \emph{correlated edges}. Each unique correlated edge is associated with a subset of \emph{Errors} associated with the corresponding \emph{ParentEdge}. The correlated pre-analysis operates as follows:
\begin{enumerate}
    \item We analyze every gate in the computation, and generate a graph with each edge containing a list of errors. Each error will be labeled with a generating gate, and some of these errors containing a list of decomposed edges. This analysis can be done without excessive duplicate processing.
    
    \item We calculate the total probability of each edge in the graph. The final edge probability $p_f$ is then approximated as simply  $p_f = p_e + p_c$. We use the same expression, but with different parameter values, to compute $p_e$ (the edge probability) and $p_c$ (the correlated edge probability). See the following Section for more details.   
\end{enumerate}

\begin{equation}
\label{one_error}
p_e = \sum_i p_i \prod_{j\neq i} (1 - p_j).
\end{equation}

One could simply add up the probability of each error associated with the edge, however at higher error rates this is inaccurate and can lead to edge probabilities above 1. Instead, it is better to group the errors associated with an edge by gate, sum the probabilities of errors associated with each gate to give a list of probabilities ${p_i}$, then approximate the edge probability $p_e$ as the probability of exactly one independent error occurring.

\subsection{Correlated reweighting and pre-matching}
\label{sec:prem}

The pre-analysis generated many edges located in space-time, each with its own probability ($p_e$), and probability of being correlated with other nearby edges ($p_c$). The probability $p_c$ of each correlated edge can be calculated using Eq.~\ref{one_error} with appropriately reduced ${p_i}$ values. The relative probability of each correlated edge is divided by the edge probability $p_e$.

We present a heuristic for choosing highly likely edges in the detection graph that will be used to reweight it. We choose edges according to the following parallel pre-matching algorithm (see Fig.~\ref{fig:prematching}):

\begin{itemize}
    \item For each detection event $e_0$, find the set of neighboring detection events $e_i$ with lowest weight connecting edges. If the set contains more than one, choose the one discovered first $e_1$ using any canonical ordering of the edges.
    \item For each detection event $e_0$ with a chosen neighboring detection event $e_1$, ask $e_1$ if $e_0$ is the chosen detection event of $e_1$. If yes, associate the two.
\end{itemize}

\begin{figure}[!t]
    \centering
    \includegraphics[width=\columnwidth]{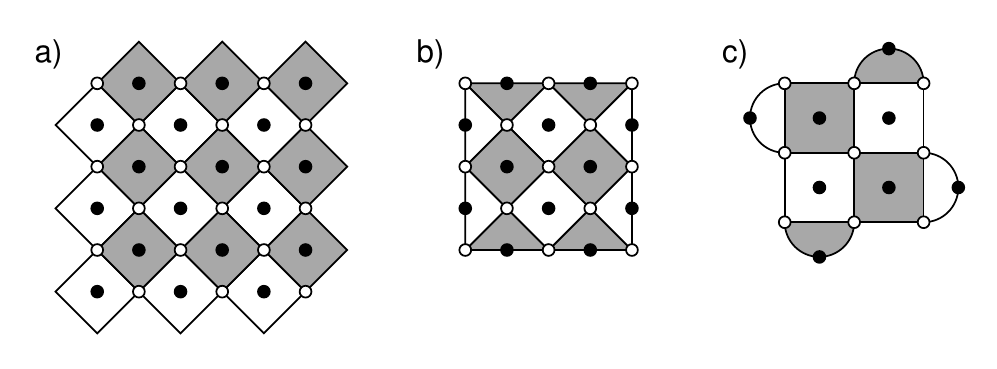}
    \caption{Distance 3 a) toric, b) unrotated, and c) rotated surface codes. Dark plaquettes represent $X$ stabilizers, light plaquettes represent $Z$ stabilizers. In all 3 cases the logical $X$ operator of interest runs from top to bottom.}
\label{sim cases}
\end{figure}

The first application of this algorithm will be called a ``virtual'' pre-matching, and will be used purely for detection graph reweighting. Given a virtual pre-matching, we can use the information about correlated edges derived in Section~\ref{sec:corr} to associate additional correlated probabilities $p_c$ with nearby edges. If more than one correlated probability is associated with a single edge, we permit the algorithm to have a race condition and only keep the last written value. The final edge probability $p_f$ is then approximated as simply  $p_f = p_e + p_c$. This in turn becomes a new weight for the edge via $w=-\ln p_f$. Additional rounds of pre-matching are optional. Decoding can then be performed with the matchings kept from pre-matching time and passed on to the general matching algorithm, or other decoder.

\section{Results}
\label{sec:simulations}

\begin{figure}[!ht]
\centering
\includegraphics[width=\columnwidth]{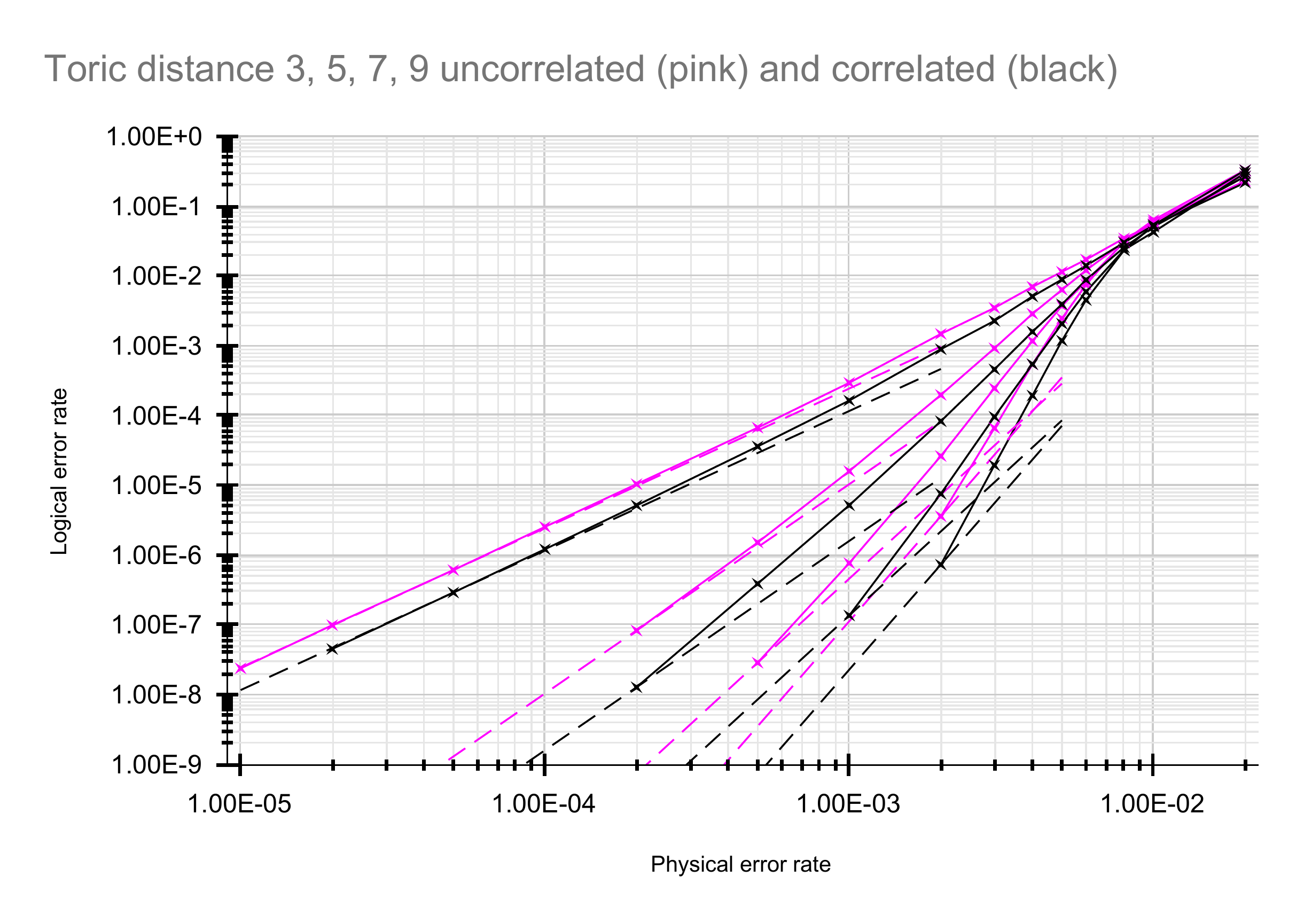}
\caption{Toric uncorrelated (pink) and correlated (black) distance 3--9 simulations. Dashed lines show $p^2$, $p^3$, $p^4$ and $p^5$ lines respectively, the asymptotic slope of each line. The fact that the data curves are much steeper than the asymptotic curves for high distances and high gate error rates shows that logical errors are suppressed at even higher powers than these in this regime.}
\label{fig:sim_1}
\end{figure}

\begin{figure}[!h]
\centering
\includegraphics[width=\columnwidth]{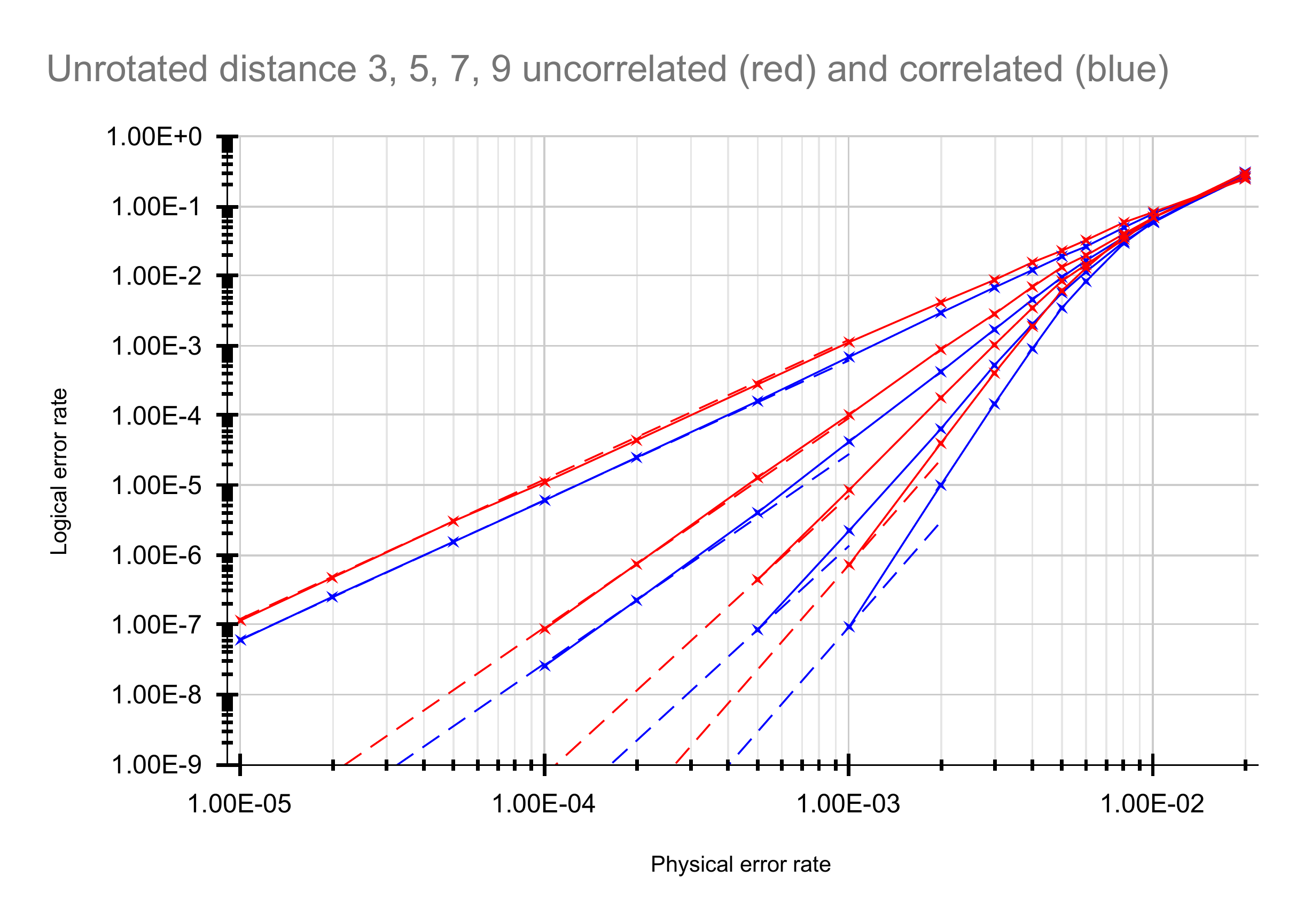}
\caption{Unrotated uncorrelated (red) and correlated (blue) distance 3--9 simulations.}
\label{fig:sim_2}
\end{figure}

\begin{figure}[!h]
\centering
\includegraphics[width=\columnwidth]{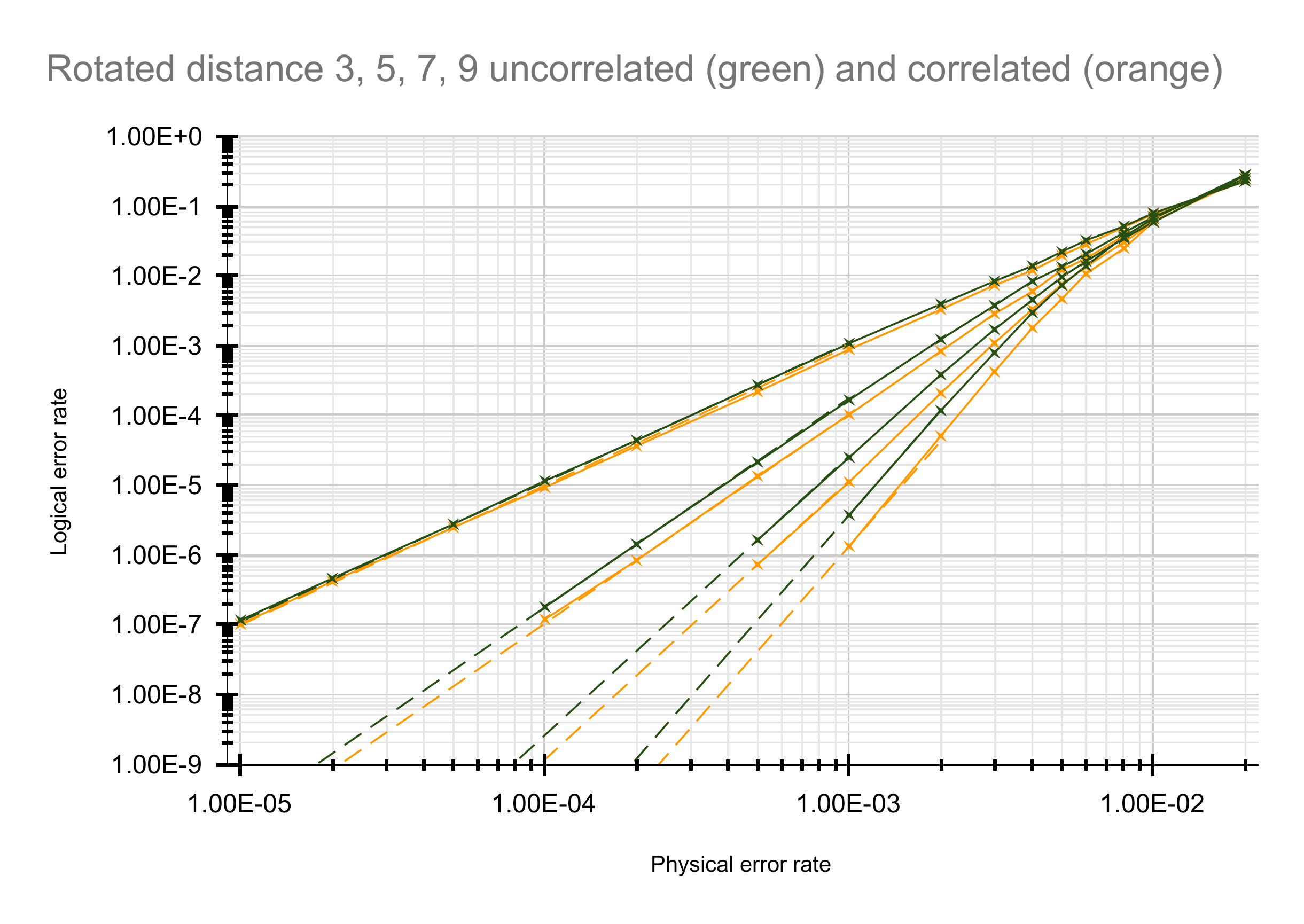}
\caption{Rotated uncorrelated (green) and correlated (orange) distance 3--9 simulations.}
\label{fig:sim_3}
\end{figure}

To illustrate the performance of our decoding algorithm, we will simulate three cases, the toric, unrotated, and rotated surface codes (Fig.~\ref{sim cases}), each with and without correlated graph re-weighting. Simulation results can be found in Figs.~\ref{fig:sim_1}--\ref{fig:sim_3}.

To the best of our knowledge, this is the first time these 3 commonly studied cases have been simulated with a single framework with results gathered in one place for comparison. We will use standard equal gate duration 8-step CNOT-based circuits as shown in Fig.~\ref{fig:de_ex}a, with all gates suffering standard depolarizing noise of equal probability $p$. Explicitly, initialization and measurement prepare and measure the wrong states with probability $p$, single-qubit gates including the identity suffer $X, Y, Z$ errors each with probability $p/3$, and CNOT gates suffer all 15 non-trivial tensor products of $I, X, Y, Z$ each with probability $p/15$.

We focus on the logical $X$ failure rate per round, and measure this by simulating a sufficiently large number of rounds $N$ such that the final probability of logical error is around 10\%, then equate this with the probability of obtaining an odd number of logical errors in $N$ and back out the failure rate per round using a sum of binomial terms.

The simulations of Fig.~\ref{fig:sim_2} are indistinguishable from those reported in \cite{Fowl13g}, with the distance 3 logical $X$ error in both cases just below $10^{-7}$ at a physical error $p$ of $10^{-5}$, and the distance 9 logical $X$ error also in both cases just below $10^{-7}$ at a physical error $p$ of $10^{-3}$.

\section{Conclusion}
\label{conclusion}

A pipeline approach is a step towards real-time decoding of the surface code. We presented a method that operates on the stream of detection events and processes the data in a sequence stages implemented in a parallelizable manner that requires no communication.

The stages of the pipeline are: reweighting, pre-matching and full-matching. The functionality of the first two stages is based on a novel analysis of correlated errors. The correlated analysis is decoder agnostic, and the graph re-weighting highly hardware compatible. One can imagine measurements from a quantum computer streaming through dedicated hardware to generate a re-weighted graph with optional partial matching that is then passed on to another decoder of the users choice.

Simulation results for the toric, rotated and unrotated surface code support the feasibility of the pipeline approach. Future work will focus on extending our results to other decoder types.

\section*{Acknowledgments}

AP was supported by Google Faculty Research Awards and a Fulbright Senior Researcher Fellowship.

\bibliographystyle{quantum}
\bibliography{__main}

\appendix

\section{Appendix: Pre-matching}
\label{sec:appendix}

This section describes the method used during the analysis of correlated errors. Pre-matching is a best effort, very fast (low complexity) decoder that is not guaranteed to have a threshold or to be useful on its own for full error-correction. However, used within a decoding pipeline, pre-matching can inform later pipeline stages about the potential matchings existing in the detection graph. Pre-matching is used for computing the graph edges which should be reweighted.

Pre-matching uses only local vertex information without explicitly including a method for achieving a global optimum low weight matching. It is a greedy algorithm that assigns and updates three types of states to a detection event: zero-prematched (ZP), half-prematched (HP) and fully-prematched (FP). State transition rules are listed in Table~\ref{tbl:st} and discussed in Section~\ref{sec:updates}. Pre-matching operates the following steps:
\begin{itemize}
    \item it initializes the states of the detection graph vertices to ZP;
    \item it iterates through all the vertices of the detection graph and updates their states based on the weights of the neighborhood edges (the neighboring vertex connected by an edge of minimum weight).
\end{itemize}

\begin{figure}[!h]
    \centering
    \includegraphics[width=\columnwidth]{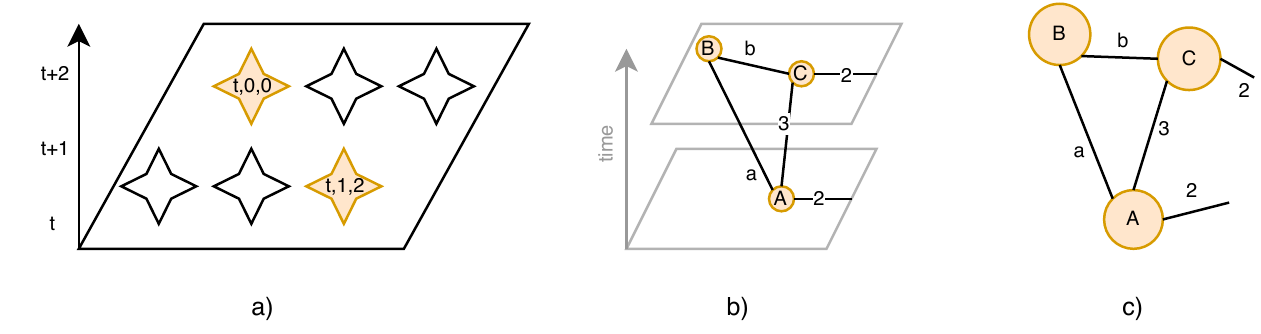}
    \caption{Pre-matching: a) Detection events have 3D coordinates, and are processed in the order of their time, row and column coordinates. For example, the first event to process is (1,0,0) and the second is (1,1,2). b) A graph of three detection events (A, B, C) is built and the weights of the edges are computed. In this example, only for the events A and C we consider the edges to the boundary of the code patch. c) The graph after eliminating the time layers visualization.}
    \label{fig:prematch1}
\end{figure}

To this end an error detection graph like the one from Fig.~\ref{fig:prematch1} is used. Each graph vertex has an associated 3D coordinate of the type $(t, i, j)$ where $(i, j)$ are coordinates of the physical qubit that generated the detection event, and $t$ is an integer indicating time. Later detection events have higher values of the time coordinate $t$. Two examples are presented in Figs.~\ref{fig:prematch2} and \ref{fig:prematch3}.  The temporal ordering of the detection events plays a role in how states are assigned and transformed.

The pre-matching algorithm has a linear complexity in the number of vertices in the detection graph. If the edges towards the neighboring detection events are not sorted based on their weights, the complexity becomes quadratic in the number of vertices, because we need to iterate through the edge list and select the one of minimum weight. The number of detection events scales quadratically with the distance $d$ of the code, and we use sorted edge lists, such that the total complexity of pre-matching is on the order of $\mathcal{O}(d^2)$.

\subsection{Pre-matching condition (PMC)}

We use the symbol $\leftrightarrow$ to indicate that two events A and B are FP, fully prematched. A full pre-match between two events A and B works in two directions: A is prematched with B, iff B is prematched with A. We call \emph{PMC} the event pre-match condition that establishes one direction of the prematching, either $\rightarrow$ or $\leftarrow$. The PMC of all the events is processed in the following order: a) increasing time $t$; b) increasing  $i$-coordinate and c) increasing $j$-coordinate.

Two events A and B are fully prematched, the $A \leftrightarrow B$ relation exists, if $A \rightarrow B$ and $A \leftarrow B$. We use both directions of the arrows ($\rightarrow$ and $\leftarrow$) because we assume a temporal ordering between the detection events A and B (cf. Table~\ref{tbl:st}). If the PMC is checked from A towards the future B then $A \rightarrow B$, otherwise if from a later B to a sooner A then $A \leftarrow B$.

A strict PMC can be formulated as $A \rightarrow B$ if B is the only low weight neighbor of A in the error detection 3D graph. The condition guarantees that the prematched events are also valid matchings from the perspective of minimum weight perfect matching. However, at higher physical error rates, the strict PMC is very seldom fulfilled, because multiple detection events of the same weight can exist at the same time in the neighborhood of another event. Thus, we introduce a relaxed PMC condition, such that more events are prematched without offering any guarantees that these form minimum weight matches.

\begin{figure}
    \centering
    \includegraphics[width=\columnwidth]{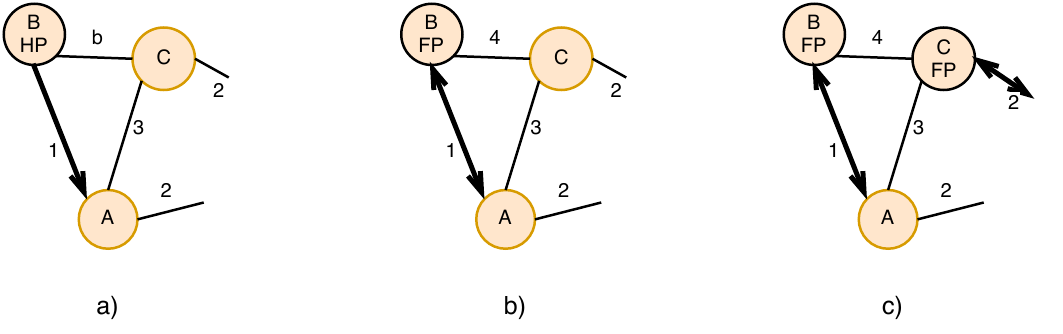}
    \caption{Advancing the state of the vertices. All vertices start in ZP (not illustrated) and the pre-match order is A, B, C. Example obtained by replacing $a=1$ and $b=4$ in Fig.~\ref{fig:prematch1}c.: a) The lowest weight edge is the one connecting A to B, such that B is marked as half prematched HP stores a reference to A; b) After pre-matching B, the lowest weight edge points to A, and the state of B is updated to full prematched FP; c) The lowest edge weight of C is the one connecting to the boundary, and C is automatically fully prematched.}
    \label{fig:prematch2}
\end{figure}

Initially, all detection events are in the state ZP. Pre-matching is performed by iterating the detection events towards the future (increasing time coordinate). Assuming that pre-matching has reached vertex $A$, half-pre-matching is performed towards the future: if the event $B$ is later than $A$, and $A \rightarrow B$ then B will be in the HP state (if it was ZP) and will store a reference to $A$ (e.g. Fig.~\ref{fig:prematch2}a). When it is the turn of B to be analysed for prematching, if $C \leftarrow B$ and C is the same as the stored reference to A, then the state of B is transformed from HP to FP.

\subsection{Vertex state updates}
\label{sec:updates}

Our pre-matching procedure updates the states of present and future detection events. If an event's state is ZP and it can be connected to the boundary, then the state is automatically updated to FP and the next detection event is considered (Section~\ref{sec:boundary}). Otherwise, the state transformation procedure from the following paragraphs is applied.

\begin{table}
    \centering

    \begin{tabular}{| c|c|c|c|}
        \multicolumn{4}{c}{coord(B) $<$ coord(A)}\\
        \hline
           \multirow{2}{*}{\textbf{A}} & \multicolumn{3}{c|}{B} \\
        \cline{2-4}
                &  ZP  &   HP   &   FP  \\
        \hline
            ZP  &   ZP  &   E   &   ZP  \\
            HP  &   FP/ZP&  E   &   E/ZP\\
            FP  &   E   &   E   &   E   \\
        \hline
    \end{tabular}

    \begin{tabular}{| c|c|c|c|}
    \multicolumn{4}{c}{coord(A) $<$ coord(B)}\\
    \hline
    \multirow{2}{*}{\textbf{B}} & \multicolumn{3}{c | }{A}\\
    \cline{2-4}
            &   ZP  &   HP          &   FP \\
    \hline
        ZP  &   HP  &    ZP$_A$     &   E \\
        HP  & E/ZP  & E/ZP$_{A,B}$  &   E \\
        FP  &   E   &       E       &   E \\
    \hline
    \end{tabular}
   
    \caption{The state transformation rules when performing pre-matching starting from detection event A. Detection event B has the lowest weight from the neighborhood of A and is fulfilling the PMC. Depending on the coordinate of B, returned by the function $coord$, one of the two state transition tables is used. When not indicated by a subscript, the state of the latest (bold) detection event is updated.}
    \label{tbl:st}
\end{table}

The pre-matching procedure does not update states of detection events from the past. Considering that A is the detection event that is processed, and that the PMC function returns B, there are two possibilities:
\begin{itemize}
    \item B has a coordinate lower than A;
    \item B has a coordinate higher than A.
\end{itemize}

\begin{figure}[t]
    \centering
    \includegraphics[width=\columnwidth]{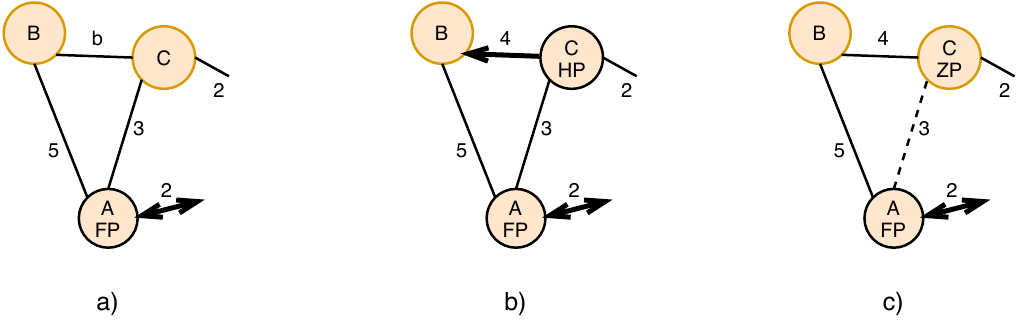}
    \caption{Reverting the state of the vertices. All vertices start in ZP (not illustrated) and the pre-match order is A, B, C. Example obtained by replacing $a=5$ and $b=4$ in Fig.~\ref{fig:prematch1}c.: a) The lowest weight edge is connecting to the boundary, and A is automatically fully prematched. b) When pre-matching B, the lowest weight edge is towards C which is half prematched and will store a reference to B. c) Because A is FP and in the neighborhood of C, the lowest weight edge (2) is not considered and the next best option is pointing towards A (3), such that C cannot be prematched and its state is reset to ZP.}
    \label{fig:prematch3}
\end{figure}

For the first situation, $coord(B) < coord(A)$ (cf. Fig.~\ref{fig:prematch3}c where C is processed and $coord(A) < coord(C)$), if B has state HP, then it is always an error, because B should have been already processed and its state is not valid. If the state of A is FP, this is impossible and always and error, because A is only now being processed. This leaves only four non-erroneous state transitions:
\begin{enumerate}
    \item If A is ZP and B is ZP, then the state of A is not changed, because B is in the past.
    \item If A is half-prematched HP and B is ZP, if the backwards reference from A points to B then both PMC directions are fulfilled and B is fully prematched and its state transitions to FP.
    \item If the reference does not point to B, then the state of A is reset to ZP. 
    \item If A is HP but the B from the past is already fully prematched, then it is an obvious error if the reference from A points to B, and otherwise the state of A is reset to ZP.
\end{enumerate}

In the second situation, $coord(A) < coord(B)$, it is always an error if A is FP. Also it is not possible for B to be FP, because it is in the future and could not have been processed by now (cf. Fig.~\ref{fig:prematch2}a where A is processed). This leaves four possible configurations:
\begin{enumerate}
    \item If both A and B are ZP, then the state of B is changed to HP and B will store a reference to A.
    \item If A is HP and the future B is still ZP, it means that the backwards PMC for A is not fulfilled and the state of A is reset to ZP.
    \item If B is HP and its reference points to A, it is definitely an error, because A is only now being processed and could have not changed the state of B to HP. Otherwise, if B is HP and A is ZP, then the state of A is kept ZP and the state of B is reset to ZP (this is a strict rule, the state of B could have been kept HP and be processed only later).
    \item If B is HP and A is HP, the states of both A and B is reset to ZP.
\end{enumerate}

The state transitions from in Table~\ref{tbl:st} are independent of the PMC. The pre-matching algorithm raises an error E each time it encounters an inconsistent (ie. incorrect) cases of vertex states. Inconsistencies are any situations where the following two rules do not hold.
\begin{enumerate}
    \item before processing a vertex, the only valid states are ZP and HP (if it has been half-matched in the past). In other words, there is an error if the current node is already fully matched without having been processed first (the third row when $coord(A) < coord(B)$, and the third column when $coord(A) > coord(B)$);
    \item after processing an vertex, the only valid states will be ZP and FP. In other words a processed vertex can be either not matched or  fully matched, and nothing in between (e.g.  HP);
\end{enumerate}

Checking all the state transitions from Table~\ref{tbl:st} might seem trivial, but pre-matching is designed to be compatible with multi-threaded (ie. parallel, distributed) stream (ie. operations in the past are not allowed) processing. In such a setting, the detection graph is a resource shared and operated by all the threads. Data consistency and correctness needs to be ensured during the concurrent access of all the threads. Checking correctness is generally challenging when implementing parallel/distributed algorithms. However, our pre-matching algorithms is designed to be easily checked for correctness, and  Table~\ref{tbl:st} is an exhaustive method of checking that parallel, multi-threaded pre-matching has been implemented and is operating correctly.

\subsection{Pre-matching with the boundary}
\label{sec:boundary}

The goal of pre-matching is to pair detection events and avoid where possible matching with the boundary (e.g. Fig.~\ref{fig:prematch3}c). Pre-matching with the boundary is avoided as, in general, locally matching with the boundary seems to be the lowest weight choice, but when considering the global sum of the weights this is often not the case.

Matching with the boundary is allowed whenever a detection event is processed and no other detection event in its neighborhood has the state FP (valid state from the past, cf. Table~\ref{tbl:st}) or HP (valid state in the future, cf. Table~\ref{tbl:st}). Considering two neighboring detection events A and B where both could be connected to the boundary we assume that such a decision will not result in a low weight global matching. For small distances, the ratio of events close to the boundary is higher than for larger distances. However, we consider that decoding should perform well for large distances and that our heuristic will result in \emph{more realistic looking} matches.

\end{document}